\documentclass[]{aa}
\usepackage{psfig}
\usepackage{txfonts}
\begin{document}

\title{An explanation for the unusual IMF in Taurus}

\author{Simon\,P.\,Goodwin, A.\,P.\,Whitworth \and D.\,Ward-Thompson}

\offprints{Simon.Goodwin@astro.cf.ac.uk}

\institute{Department of Physics \& Astronomy, Cardiff University, PO Box 913,
5 The Parade, Cardiff, CF24 3YB, UK}


\abstract{
In comparison with other well studied star formation regions, Taurus 
is unusual in several respects. (i) Its stellar initial mass function 
(IMF) peaks at relatively high mass ($\sim 0.8 M_\odot$), but contains 
very few stars much more massive than $1 M_\odot$, and is relatively 
deficient in brown dwarfs. (ii) It has a higher binary fraction, particularly 
at large separations. (iii) Its core mass function is strongly peaked at a few 
$M_\odot$, and the cores have extended envelopes and relatively low levels 
of turbulence.

\hspace{0.25cm} We present here the results of an ensemble of hydrodynamic 
simulations which suggest that the unusual stellar IMF in Taurus is a 
direct consequence of the unusual properties of the cores there. By 
following the collapse and fragmentation 
of cores having properties typical of Taurus, we find that roughly 50\% of 
the objects formed in a core, predominantly the low-mass ones, are ejected 
from the core to form a population of low-mass stars and brown dwarfs with 
a flat mass function. The remaining objects form multiple systems within 
the core, accreting until their masses approach $1 M_\odot$; this produces a 
population of intermediate-mass stars whose mass function peaks at $\sim 
0.8 M_\odot$. Together these two populations reproduce the IMF in Taurus 
very well. This demonstrates, for the first time, a direct causal link 
between the core mass function and the stellar IMF in a star formation 
region.

\keywords{Stars: formation -- ISM: clouds-structure -- Methods: numerical}
}

\maketitle

\section{Introduction}

Stars form within dense molecular cores (e.g. Andr\'{e} et al., 2000), and 
the densest and most centrally condensed cores, i.e. those closest to forming 
stars, are known as prestellar cores (Ward-Thompson et al., 1994). In Ophiuchus 
(Motte, Andr\'e \& Neri, 1998), Serpens (Testi \& Sargent, 1998) and 
Orion (Motte et al., 2001), the mass function of 
pre-stellar cores is remarkably similar to the IMF for field stars and 
clustered star formation regions. This suggests that there is a simple 
mapping from the core mass function into the stellar IMF, with the mean 
masses of the stars forming within a core being proportional to the core 
mass. However, the details of how this works are still uncertain.

In this paper we present numerical simulations of core collapse and 
fragmentation which demonstrate a causal link between the core mass
function and the stellar IMF in the Taurus star formation region. 
We choose to study the Taurus region, because it is nearby and extended 
on the sky, and has therefore been studied in detail. Molecular-line 
mapping has yielded estimates of the Taurus core mass function (Onishi et al., 
2002), and deep surveys of its stellar content have revealed the Taurus 
IMF down to the deuterium burning limit (Luhman et al., 2003a).

\section{Observational background}

Determining and explaining the stellar initial mass function (IMF) -- i.e. 
the number of stars, ${\cal N}_M\,dM$, forming with mass in the interval 
$(M,M+dM)$ -- is one of the longest standing problems in the study of star 
formation.

Salpeter (1955) first investigated the IMF for field stars in the solar 
neighbourhood and found a power-law fit, ${\cal N}_M \propto M^{-\alpha}$ 
with $\alpha \simeq 2.35$, for stars in the mass range $(0.4 M_\odot < M 
< 10 M_\odot)$. Subsequent work (e.g. Miller \& 
Scalo, 1979; Scalo, 1986; Kroupa, 2002) has extended the range to both 
lower and higher masses. In the field, Kroupa (2002) finds that Salpeter's 
result is still a good approximation for intermediate and high masses 
($M > 0.5 M_\odot$), but at lower masses the stellar IMF flattens (i.e. 
$\alpha$ is smaller). Specifically, $\alpha \sim 1.3 \pm 0.5$ for low-mass 
stars, $0.08 M_\odot < M < 0.5 M_\odot$; and $\alpha \sim 0.3 \pm 0.7$ 
for brown dwarfs, $0.01 M_\odot < M < 0.08 M_\odot$.

In individual star formation regions, there appears to be some variance 
in the stellar IMF, but in regions of {\it clustered} star formation 
this variance is small. For example, Muench et al. (2002) and Luhman 
et al. (2003b) find stellar IMFs broadly similar to Kroupa's field 
star IMF in -- respectively -- Orion and IC348.

\begin{figure*}
\centerline{\psfig{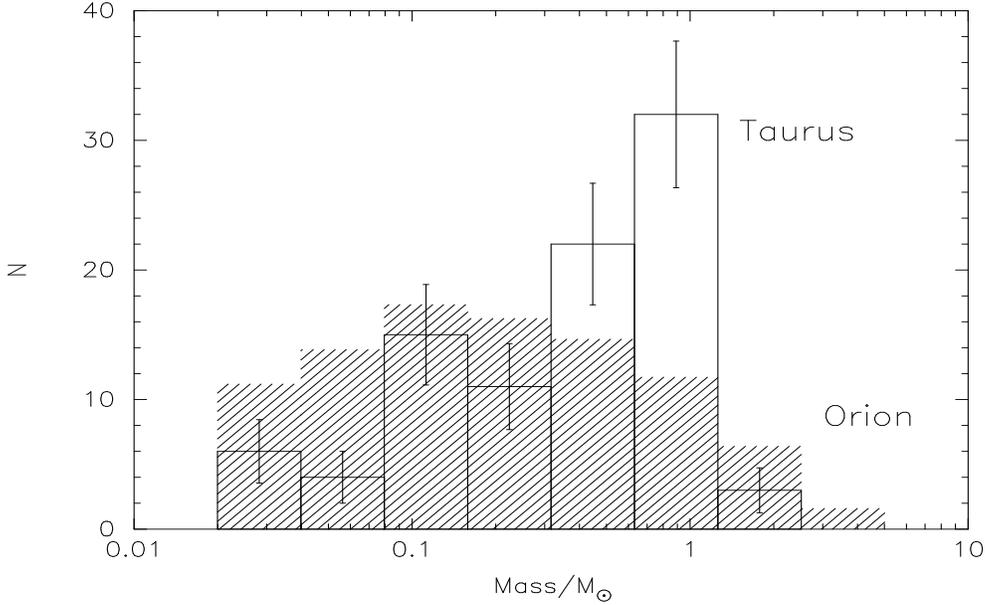}}
\caption{A plot of number of stars, $N$, versus logarithmic stellar mass, 
$\ell og_{10}[M]$, comparing the IMF in Taurus (solid-line histogram, 
adapted from Luhman et al., 2003a) with that in Orion (shaded histogram, 
adapted from Muench et al., 2002). The Orion IMF is normalised to have
the same number of stars as the Taurus IMF. The two IMFs are clearly 
different.}
\label{fig:compimf}
\end{figure*}

However, in Taurus, which is a region of {\it distributed} star formation, 
the stellar IMF is markedly different. Figure~\ref{fig:compimf} compares 
the IMF of Taurus (solid-line histogram, adapted from Luhman et al., 
2003a) with 
that of Orion (shaded histogram, adapted from Muench et al., 2002). We 
note the following points from Figure~\ref{fig:compimf}: 
\begin{itemize}
\item The Taurus IMF peaks at a higher mass, $\sim 0.8 M_\odot$ (as compared 
with $\sim 0.2 M_\odot$ in Orion).
\item The Taurus IMF contains very few stars more massive than $1 M_\odot$ 
(whereas Orion contains several stars above $10 M_\odot$).
\item The Taurus IMF contains many fewer brown dwarfs per star  
than Orion (less than half as many)
\item At the low-mass end the Taurus IMF is approximately flat.
\end{itemize}

There is further evidence that the pattern of 
star formation in Taurus is different from that in regions of clustered 
star formation. 
\begin{itemize} 
\item There is a higher binary fraction in Taurus than in Orion, 
particularly at large separations. 
\item The prestellar cores in Taurus (Onishi et al., 2002) have a narrower mass 
function than those in Orion (Motte et al., 2001) or Ophiuchus (Motte et al., 
1998); most cores in Taurus have $M \sim 5 \pm 3 M_\odot$. 
\item The prestellar cores in Taurus are on average more extended and less 
turbulent than those in Orion and Ophiuchus.
\end{itemize}

\section{Modelling the collapse and fragmentation of pre-stellar cores in Taurus}

\subsection{Initial conditions and constitutive physics}

In view of the strongly peaked core mass function in Taurus, we 
consider a single core mass, $M = 5 M_\odot$. The initial conditions 
are chosen to mimic the cores observed in Taurus (Ward-Thompson et al., 
1994; Ward-Thompson, Motte \& Andr\'e, 1999; Jijina, Myers \& Adams, 1999; 
Ward-Thompson, Andr\'e \& Kirk, 2002; Kirk, Ward-Thompson \& Andr\'e, 
2003). The core density profile is initially
\begin{equation}
\rho = \rho_{\rm kernel}\,\left[ 1 \,+\, 
\left( \frac{r}{r_{\rm kernel}} \right)^2 \right]^{-2} \,,
\;\; r \, < \, r_{\rm boundary}
\end{equation}
with $\rho_{\rm kernel} = 3 \times 10^{-18}\,{\rm g}\,{\rm cm}^{-3}$ 
(corresponding to $n_{{\rm H}_2} \simeq 7 \times 10^5\,{\rm cm}^{-3}$), 
$r_{\rm kernel} = 5,000\,{\rm AU}$ and $r_{\rm boundary} = 50,000\,{\rm AU}$. 

The isothermal sound speed in a core is initially $c_0 = 0.19\,{\rm km}\,
{\rm s}^{-1}$ (corresponding to molecular gas at $10\,{\rm K}$), and so 
the initial Jeans mass in the centre of a core is $\sim 0.8 M_\odot$, and 
the initial ratio of thermal to gravitational potential energy is 
$U_{\rm thermal}/|\Omega| = 0.45$. In order to capture the switch from 
isothermality to adiabaticity which is expected to occur as the density 
rises above $\rho_{\rm crit} \sim 10^{-13}\,{\rm g}\,{\rm cm}^{-3}$ 
(eg. Tohline, 1982; Masunaga \& Inutsuka, 2000), we use the equation of state
\begin{eqnarray}\label{EofS}
P \;=\; \rho\,c_0^2\,\left[ 1 \,+\, \left( \frac{\rho}{\rho_{\rm crit}} 
\right)^{2/3} \right] \,.
\end{eqnarray}

A divergence-free Gaussian random velocity field is superimposed on each 
core to simulate turbulence. The power spectrum of the turbulence is 
$P(k) \propto k^{-4}$. Fifty realizations of this representative core are 
treated. The only difference between the different realizations is the 
random number seed used to generate the turbulent velocity field, and the 
initial ratio of turbulent to gravitational potential energy, 
$U_{\rm turbulent} / |\Omega|$. In order to match the levels of turbulence 
measured in Taurus, twenty cores have $U_{\rm turbulent} / |\Omega| = 0.05$, 
twenty cores have $U_{\rm turbulent} / |\Omega| = 0.10$, and ten cores have 
$U_{\rm turbulent} / |\Omega| = 0.20\,$.

\begin{figure*}
\centerline{\psfig{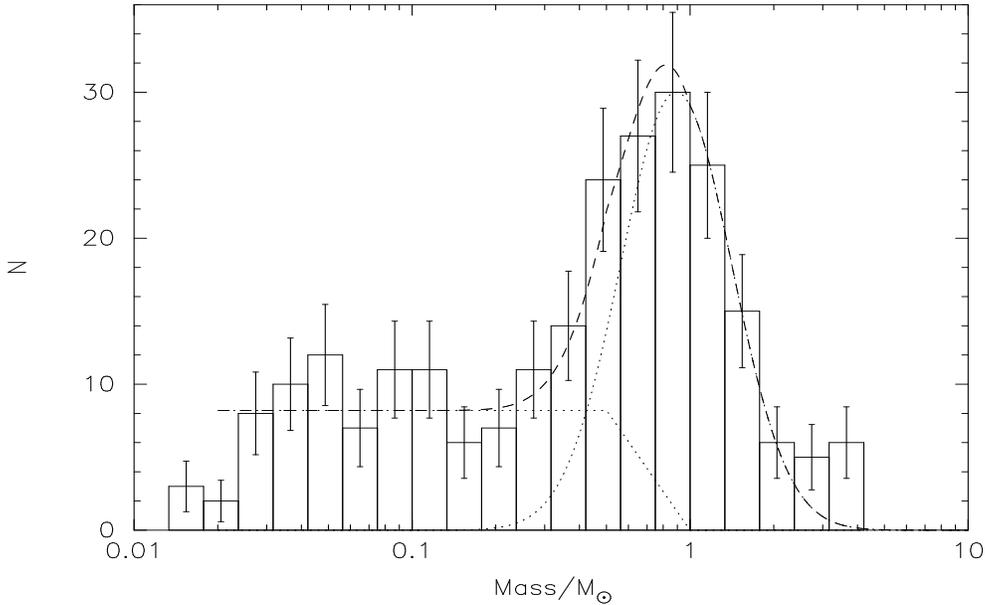}}
\caption{The stellar IMF produced from an ensemble of fifty 
core collapse simulations is shown as a solid-line histogram. This histogram 
is well-fitted by the sum (dashed line) of two distributions (dotted lines): 
a log-normal distribution with mean $\ell og_{10}[M] = - 0.05$ and standard 
deviation $\sigma_{ \ell og_{10}[M]} = 0.02$ {\it plus} a flat distribution 
below $0.5 M_{\odot}$ The stars making up the log-normal distribution 
generally tend to remain in multiple systems, whilst those making up the 
flat distribution are generally objects that have been ejected from multiple 
systems.}
\label{fig:theoryIMF}
\end{figure*}

\subsection{Numerical method}

The simulations are performed with a standard SPH code (e.g. Lucy, 1977; 
Gingold \& Monaghan, 1977; Monaghan, 1992). For details of 
the code, see Turner et al. (1995) and Goodwin et al. (2004a,b). It uses 
variable smoothing lengths $h$, the M4 smoothing kernel (Monaghan \& 
Lattanzio, 1985), and ${\cal N}_{\rm neib} = 50 \pm 5$ neighbours. 
The artificial viscosity prescription of Lattanzio et al. (1986) is 
used with $\alpha_{\rm visc} = 1$ and $\beta_{\rm visc} = 2$. An octal 
tree (Barnes \& Hut, 1986) is used to construct neighbour lists and to 
calculate gravitational accelerations, which are kernel softened using 
$h$. If the density of particle $i$ rises above $100\,\rho_{\rm crit}$, 
and the particles within $5\,{\rm AU}$ of particle $i$ are gravitationally 
bound, they are replaced with a sink particle of radius $5\,{\rm AU}$ 
(Bate, Bonnell \& Price, 1995). 

Most simulations are made with ${\cal N}_{\rm tot} = 25,000$ particles, 
but some have been repeated with ${\cal N}_{\rm tot} = 50,000$ particles 
in order to test convergence. In no case is there a significant difference 
in the evolution using larger ${\cal N}_{\rm tot}$. With ${\cal N}_{\rm tot} 
= 25,000$, the minimum mass which can be resolved is $M_{\rm res} \simeq 
0.01 M_\odot$. With the equation of state given by Eqn. (\ref{EofS}), the 
minimum Jeans mass is $M_{\rm min} \simeq 0.01 M_\odot$. Therefore the Jeans 
condition (Bate \& Burkert, 1997; Whitworth, 1998) is satisfied.

\subsection{Results}

We have previously shown (Goodwin et al., 2004a,b) that even cores with 
very low levels of turbulence can fragment to form multiple systems. In 
the ensemble of fifty simulations of Taurus-like cores reported here, each 
core forms between 1 and 9 objects, with an average of 4.5 objects. 
Details of the individual simulations are given in Goodwin et al. (2004a,b). 
Typically the lower-mass objects are ejected from the core in a period 
of rapid dynamical evolution (cf. Sterzik \& Durisen, 2002), and the 
higher-mass objects remain in the core and form multiple systems. The ejected 
objects are lower-mass because they are ejected before they can grow much 
by accretion (cf. Reipurth \& Clarke, 2001; Bate, Bonnell \& Bromm, 2002; 
Delgado-Donate et al., 2003, 2004; Goodwin et al., 2004a,b). The 
objects remaining in the core are higher-mass because they continue 
to accrete (cf. Bonnell et al., 2001).

Figure~\ref{fig:theoryIMF} shows as a histogram the mass function of 
objects produced by this ensemble of runs. The histogram is well fitted 
by the sum (dashed line) of two distributions (dotted lines): a narrow 
log-normal distribution, with mean $\left< \ell og_{10}[M] \right> = - 0.05$ 
and standard deviation $\sigma_{ \ell og_{10}[M]} = 0.02$; and a flat 
distribution below $0.5 M_{\odot}$. The stars contributing to the 
log-normal distribution are mainly those that remain in the core and 
end up in multiple systems. The stars contributing to the flat 
distribution are mainly those ejected from the core.

The low-mass objects have a flat mass distribution because their ejection 
is a stochastic process. In any individual core there are only a few 
low-mass objects, so the ejection dynamics cannot be very selective about 
which object is ejected (i.e. ejection is only weakly dependent on mass). 
The mass of an ejected object depends mainly on how long after formation 
it is ejected. If ejection occurs early, the object has had little time 
to grow and is likely to be a brown dwarf. If ejection occurs later, 
the object may have grown to become a low-mass star.

\begin{figure*}
\centerline{\psfig{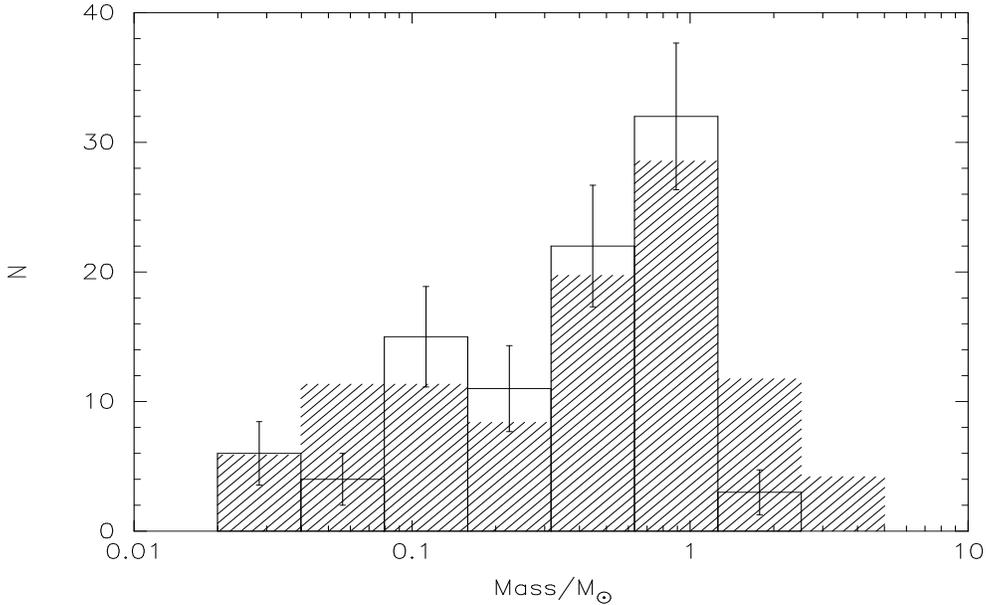}}
\caption{The solid-line histogram shows the observed IMF of Taurus due to 
Luhman et al. (2003a) (as in Figure 1). The shaded histogram shows the
model IMF from Figure 2, normalised and rebinned to match the binning of 
Luhman et al. (2003a). The model IMF was produced from an ensemble of simulations
which starts with the observed core mass distribution in Taurus and initial
conditions for individual cores (density profile, level of turbulence, etc.) 
based on the physical conditions observed in pre-stellar cores in Taurus. 
The two histograms have been normalised by minimising $\chi^2$. Agreement 
between the model and the observations is good (see text for discussion).}
\label{fig:allimfs}
\end{figure*}

\section{Comparison of simulations with observations}

Figure~\ref{fig:allimfs} shows a comparison between the observed IMF of 
Taurus from Figure 1 (solid line histogram) and the model IMF from 
Figure 2, normalised and rebinned to the same binning as the 
observations (shaded histogram). The two histograms are very similar. 
A Kolmogorov-Smirnov test estimates the probability that the underlying 
distributions are different at less than 10\%. A $\chi^2$ test accepts 
the hypothesis that they are the same at more than 90\% confidence.

The main discrepancy between the histograms is at high masses, where 
the simulations appear to produce too many stars above $1 M_\odot$. This 
may be due to the simulations not including feedback. Bipolar outflows 
from young stars would reduce accretion onto the more massive stars, 
and might thereby bring the histograms into better agreement. We are 
currently repeating a subset of the simulations with a more 
sophisticated sink model which includes feedback due to bipolar 
outflows (Boyd et al., in preparation), in order to test this hypothesis.

Brice\~{n}o et al. (2002) report that the spatial distributions of stars
and brown dwarfs in Taurus are very similar, and this would appear to be 
evidence against the hypothesis that brown dwarfs are formed by ejection 
from star forming cores (e.g. Reipurth \& Clarke, 2001; see also Kroupa \& 
Bouvier, 2003, for a discussion in the specific context of Taurus). A similar 
situation is seen in Chamaeleon I by L\'opez Mart\'i et al. (2003).

However, in our simulations the ejected objects comprise both brown dwarfs 
and low-mass stars, with almost twice as many low-mass stars ($0.08 M_\odot 
< M < 0.5 M_\odot$) as brown dwarfs ($M < 0.08 M_\odot$). Moreover, the 
ejection speed is almost independent of mass ($\sim 1 \, - \, 2\,
{\rm km}\,{\rm s}^{-1}$), and therefore the distributions and kinematics 
of brown dwarfs and low-mass stars are very similar, as observed.

Moreover, any systematic difference between the velocity dispersions of 
objects of different mass is unlikely to produce significant segregation 
on timescales less than 10 Myr, for the following reasons. First, the 
star-forming cores in Taurus are widely spaced, and so there is no single centre 
from which to measure a putative diaspora of brown dwarfs or low-mass stars. 
Second, the individual star-forming cores have different velocities; from 
Onishi et al. (2002), the inter-core velocity dispersion is 
$\sim \pm \, 1 \, {\rm km}\,{\rm s}^{-1}$. Third, some of the older, smaller star-forming cores may already have dispersed.

\section{Conclusions}

Taurus has an unusual stellar initial mass function and an unusual core 
mass function. We have modelled the hydrodynamical evolution of an ensemble 
of cores with masses based on the Taurus core mass function and 
levels of turbulence based on those observed in Taurus. We find that
the unusual stellar IMF in Taurus can be explained as a direct result 
of the unusual core mass function and intrinsic core properties in Taurus.

In each core an initially unstable, multiple system forms, with 
between 2 and 9 members. Typically, 2 or 3 objects are ejected 
before they can accrete a significant amount of material. These 
ejected low-mass stars and brown dwarfs constitute 
the flat, low-mass `tail' of the Taurus IMF. There are almost twice as many 
low-mass stars ($0.08 M_\odot < M < 0.5 M_\odot$) as brown dwarfs ($M < 0.08 
M_\odot$). The ejection velocities ($\sim 1\,-\,2\,{\rm km}\,{\rm s}^{-1}$) are 
essentially independent of mass, so the low-mass stars are as dispersed as 
the brown dwarfs and there is no significant mass segregation.

The remaining objects stay near the centre of the core and continue 
to accrete until their masses are $\sim 0.8 M_{\odot}$, by which 
stage there is not much mass left to accrete. These more massive 
stars constitute the Gaussian peak centred at $\sim 0.8 M_\odot$ 
in the Taurus IMF. Almost all of them are in binary or triple 
systems. Thus the simulated cluster of cores has an IMF very similar to 
that of Taurus, viz. a narrow Gaussian peak at $\sim 0.8 M_\odot$, 
and a flat tail at lower masses, extending into the brown dwarf
regime.

Further support for the hypothesis that the form of the IMF in 
Taurus is a direct result of the typical masses of the cores in 
Taurus comes from Goodwin et al. (in preparation) who 
find that a core with low levels of turbulence typically forms a number of
objects approximately equal to the number of initial Jeans masses 
in the core (the
initial Jeans mass being roughly $1 M_{\odot}$).  Thus lower-mass
cores form fewer objects and are unable to populate the tail of 
the IMF as they do not eject significant numbers of brown dwarfs and
low-mass stars.  Higher-mass cores over-populate the tail through
the ejection of more brown dwarfs and low-mass stars.  In addition the
stars that remain within a core have a larger resovoir of gas
to accrete and become larger than the observed $0.8 M_{\odot}$ peak.

The agreement between the two histograms of Figure 3 (observation and 
theory) is remarkably close. Hence we believe we have shown, for the 
first time, a direct causal link between the core mass function of 
a star-forming region and the stellar IMF produced in that region.

\begin{acknowledgements}

Thanks to the referee, Andi Burkert, for his useful comments.  
SPG acknowledges support of PPARC grant PPA/G/S/1998/00623 and is now a
UKAFF Fellow.

\end{acknowledgements}

\end{document}